\documentclass{article}
\usepackage{cite,amsmath,amssymb,color,tikz,fullpage,mdframed,subcaption}
\usepackage[colorlinks=true,citecolor=blue,hyperindex=true,draft=false]{hyperref}

\makeatletter
\def\endthebibliography{%
  \def\@noitemerr{\@latex@warning{Empty `thebibliography' environment}}%
  \endlist
}
\makeatother

\newmdtheoremenv[linecolor=gray,leftmargin=20,rightmargin=20,backgroundcolor=gray!40,innertopmargin=0pt,ntheorem]{myprop}{}[section]

\begin{document}

\title{On Optimality in ROVir}

\author{Justin P. Haldar}

\maketitle

We recently published an approach named ROVir (Region-Optimized Virtual coils \cite{kim2021})  that uses the beamforming capabilities of a multichannel magnetic resonance imaging (MRI) receiver array to achieve coil compression (reducing an original set of receiver channels into a much smaller number of virtual channels for the purposes of dimensionality reduction), while simultaneously preserving the MRI signal from desired spatial regions and suppressing the MRI signal arising from unwanted spatial regions.  The original ROVir procedure is computationally-simple to implement (involving just a single small generalized eigendecomposition), and its signal-suppression capabilities have  proven useful  in an increasingly wide range of MRI applications \cite{kim2021,lam2022,kim2023,schauman2022,iyer2022,iyer2023,schar2023,brackenier2023, piek2023, sorland2023,jin2023,borisch2023 }. 

Our original paper \cite{kim2021} made the following claim about theoretical optimality:
\begin{mdframed}[linecolor=gray!40,leftmargin=20,rightmargin=20,backgroundcolor=gray!40]
 ``[I]t can be shown that projecting the data onto the $N_v$-dimensional subspace spanned by the top-$N_v$ generalized eigenvectors $\mathbf{w}_1,\ldots,\mathbf{w}_{N_v}$ is optimal in the sense that it maximizes the ratio between the retained signal energy and the retained interference energy among all possible $N_v$-dimensional subspace projections \cite{koles1994}.''
\end{mdframed}
However, we did not formally present the details of the  $N_v$-dimensional optimization problem we were alluding to, nor did we provide a deep discussion of some of the mathematical nuances.  Interestingly, it turns out that the quoted claim is (1) always true  for specific choices of $N_v$ and (2) always true for all choices of $N_v$ with a specific notion of orthogonal subspace projection, but there can be choices of $N_v$ and definitions of subspace projection for which the top-$N_v$  generalized eigenvectors may be suboptimal \cite{borloz2011}.   In such cases, it can be possible to improve the ratio between the retained signal energy and the retained interference energy by modifying the way that the vectors $\mathbf{w}_1,\ldots,\mathbf{w}_{N_v}$ are chosen.  The purpose of this write-up is to elaborate on these mathematical details and demonstrate the types of improvements that are possible using alternative ROVir calculations.  This discussion is largely academic, with implications that we suspect will be minor for practical applications -- we have only observed small improvements to ROVir performance  in the cases we have tried,  and it would have been safe in these cases to still use the simpler original calculation procedure with negligible practical impact on the final imaging results. 

\section{Problem Setup and Optimality: The $N_v=1$ Case}
For the sake of brevity, this write-up will focus on the abstract mathematical aspects of ROVir, while largely ignoring the MRI-specific details.  Suppose we are given Hermitian positive semidefinite matrices $\mathbf{A},\mathbf{B}\in \mathbb{C}^{N_c\times N_c}$, where $\mathbf{B}$ is assumed to have full rank (i.e., it is strictly positive definite).  ROVir involves finding weight vectors $\mathbf{w} \in \mathbb{C}^{N_c}$ that maximize a signal-to-interference ratio (SIR) metric, and the original ROVir paper started by considering the  optimization problem with $N_v=1$:
\begin{equation}
 \hat{\mathbf{w}} = \arg\max_{\mathbf{w} \in \mathbb{C}^{N_c}} SIR(\mathbf{w}), \text{ with } SIR(\mathbf{w}) \triangleq \frac{\mathbf{w}^H\mathbf{A}\mathbf{w}}{\mathbf{w}^H\mathbf{B}\mathbf{w}}.\label{eq:1d}
\end{equation}
This optimization problem takes the form of a generalized Rayleigh quotient, and has an optimal solution that is easily calculated.  Specifically, if $\{\lambda_j\}_{j=1}^{N_c}$ and $\{\mathbf{w}_j\}_{j=1}^{N_c}$ respectively denote the generalized eigenvalues and generalized eigenvectors for the  pair of matrices $\mathbf{A}$ and $\mathbf{B}$ such that $\mathbf{A}\mathbf{w}_j = \lambda_j \mathbf{B}\mathbf{w}_j$, and if we additionally assume that the generalized eigenvalues are arranged in descending order such that $\lambda_1 \geq \lambda_2 \geq \cdots \geq \lambda_{N_c}$, then the optimal $\hat{\mathbf{w}}$ is given by $\hat{\mathbf{w}} = \mathbf{w}_1$.  (While the proof of this fact is classical, we will rederive this result below because of the intuition it can bring to the higher-dimensional optimization problem).  Conversely, if we had instead been interested in minimizing the SIR, then the optimal  $\hat{\mathbf{w}}$ would be given by $\hat{\mathbf{w}} = \mathbf{w}_{N_c}$.

\subsection{Derivation of Optimality for $N_v=1$}
Consider the optimization problem in Eq.~\eqref{eq:1d}.  Since $\mathbf{B}$ is assumed to be positive definite, we can take its Cholesky decomposition $\mathbf{B} = \mathbf{L}\mathbf{L}^H$.  Then we can consider the change of variables $\mathbf{y} = \mathbf{L}^H\mathbf{w}$, which allows us to reformulate the problem as
\begin{equation}
 \hat{\mathbf{y}} = \arg\max_{\mathbf{y} \in \mathbb{C}^{N_c}} \frac{\mathbf{y}^H\mathbf{L}^{-1}\mathbf{A}\mathbf{L}^{-H}\mathbf{y}}{\mathbf{y}^H\mathbf{y}} = \arg\max_{\mathbf{y} \in \mathbb{C}^{N_c}} \frac{\mathbf{y}^H\mathbf{L}^{-1}\mathbf{A}\mathbf{L}^{-H}\mathbf{y}}{\|\mathbf{y}\|_2^2},
\end{equation} 
with $\hat{\mathbf{w}} = \mathbf{L}^{-H}\hat{\mathbf{y}}$.  This new optimization problem is scale invariant (the cost function value does not change under nonzero scaling of the vector $\mathbf{y}$), so we can take $\|\mathbf{y}\|_2 = \gamma >0 $ without loss of generality, which leads to the simplification that 
\begin{equation}
 \hat{\mathbf{y}} = \arg\max_{\mathbf{y} \in \mathbb{C}^{N_c }} \mathbf{y}^H\mathbf{L}^{-1}\mathbf{A}\mathbf{L}^{-H}\mathbf{y} \text{ s.t. } \|\mathbf{y}\|_2=\gamma ,
\end{equation} 
which is maximized by taking $\hat{\mathbf{y}}$ equal to the $\gamma$-normalized eigenvector of $\mathbf{L}^{-1}\mathbf{A}\mathbf{L}^{-H}$ corresponding to the largest eigenvector.  Specifically, we must have that
\begin{equation}
 \mathbf{L}^{-1}\mathbf{A}\mathbf{L}^{-H} \hat{\mathbf{y}} = \lambda_1 \hat{\mathbf{y}}.
\end{equation}
This expression could be used for calculation, though is unnecessarily cumbersome because it requires  a Cholesky decomposition.  A simplification occurs if we use the fact that $\hat{\mathbf{y}} = \mathbf{L}^H\hat{\mathbf{w}}$, which leads to
\begin{equation}
\begin{split}
 &\mathbf{L}^{-1}\mathbf{A}\mathbf{L}^{-H} \mathbf{L}^H\hat{\mathbf{w}} = \lambda_1 \mathbf{L}^H\hat{\mathbf{w}} \\
 \iff & \mathbf{L}^{-H}\mathbf{L}^{-1}\mathbf{A} \hat{\mathbf{w}} = \lambda_1 \hat{\mathbf{w}}\\
 \iff & \mathbf{B}^{-1}\mathbf{A} \hat{\mathbf{w}} = \lambda_1 \hat{\mathbf{w}},
 \end{split}
\end{equation}
which tells us that $\hat{\mathbf{w}}$ is the eigenvector of $\mathbf{B}^{-1}\mathbf{A}$ corresponding to its largest eigenvalue (and that it is not actually necessary to perform a Cholesky decomposition).  Note that we can also observe from this expression that $\mathbf{A}\hat{\mathbf{w}} = \lambda_1 \mathbf{B}\hat{\mathbf{w}}$, implying that $\hat{\mathbf{w}}$ must be the generalized eigenvector corresponding to the largest generalized eigenvector of the matrices $\mathbf{A}$ and $\mathbf{B}$, such that $\hat{\mathbf{w}} = \mathbf{w}_1$ as already described.  The choice of the scalar $\gamma$ can be made such that we obtain a unit-norm vector $\hat{\mathbf{w}}$. 

\section{Higher-Dimensional Optimization ($N_v>1$)}
The original ROVir paper did not clearly define the SIR-optimization problem in higher dimensions ($N_v>1$),  and there are different ways it could be defined that will lead to potentially different optimal solutions.  One formulation would involve solving
\begin{equation}
 \hat{\mathbf{Y}} = \arg\max_{\mathbf{Y} \in \mathbb{C}^{N_c\times N_v}} \frac{\sum_{i=1}^{N_v} \mathbf{y}_i^H\mathbf{L}^{-1}\mathbf{A}\mathbf{L}^{-H}\mathbf{y}_i}{\sum_{i=1}^{N_v} \mathbf{y}_i^H\mathbf{y}_i} \text{ s.t. } \mathbf{Y}^H\mathbf{Y}=\mathbf{I} \text{ with } \hat{\mathbf{w}}_i = \mathbf{L}^{-H}\hat{\mathbf{y}}_i \text{ for } i=1,\ldots,N_v.\label{eq:orig}
\end{equation} 
This formulation is based on similar principles to those used in the $N_v=1$ case above, where a change of variables has been used to simplify the optimization problem.  In this case, the optimal  $\hat{\mathbf{w}}_i$ are indeed given by the top-$N_v$ generalized eigenvectors of the matrices $\mathbf{A}$ and $\mathbf{B}$. However, this approach is associated with $\mathbf{L}$-transformed notions of inner product and orthogonality, which may seem somewhat abstract and difficult to interpret intuitively.  This transformation has an effect on the way that orthogonal projections are calculated, which can influence the characteristics of the optimal solution when $N_v>1$.  

In what follows, we will consider a different version of the optimization problem that relies on a non-transformed SIR formulation.  When computing projections onto the span of $N_v$ vectors in higher dimensions, expressions are simplified if the vectors are orthonormalized, and we will adopt this approach for the rest of our derivations.  For example, if we have orthonormal vectors $\mathbf{w}_1,\ldots,\mathbf{w}_{N_v}$ (with the conventional Euclidean notion of orthogonality), then if we project the data onto the subspace spanned by these vectors (using the conventional Euclidean orthogonal projection), the ratio between the total retained signal energy and the total retained interference energy (using the conventional Euclidean definition of energy) would be simply given by
\begin{equation}
  SIR(\mathbf{w}_1,\mathbf{w}_2,\ldots,\mathbf{w}_{N_v}) \triangleq \frac{\sum_{i=1}^{N_v} \mathbf{w}_i^H\mathbf{A}\mathbf{w}_i}{\sum_{i=1}^{N_v}\mathbf{w}_i^H\mathbf{B}\mathbf{w}_i}.\label{eq:sir}
\end{equation}
Interestingly, while finding vectors that maximize this ratio is straightforward in some situations (e.g., for $\mathbf{B}=\mathbf{I}$ regardless of the choice of $N_v$; for $N_v=1$ as derived in the previous subsection; and also for $N_v=N_c-1$ and $N_v=N_c$), it can be substantially more challenging in other situations.  For example, this optimization problem has been studied in a different context by Borloz and Xerri \cite{borloz2011}, where it was shown that for general $N_v$ and general $\mathbf{B}$, there are situations for which it is suboptimal to choose $\mathbf{w}_1,\mathbf{w}_2,\ldots,\mathbf{w}_{N_v}$ to have the same span as the top-$N_v$ generalized eigenvectors of $\mathbf{A}$ and $\mathbf{B}$.  In addition, Borloz and Xerri also argue that, aside from certain special cases, there are usually not simple (e.g., recursive/greedy) algorithms that will yield the globally-optimal choice of $\mathbf{w}_1,\mathbf{w}_2,\ldots,\mathbf{w}_{N_v}$.

To illustrate the suboptimality of choosing $\mathbf{w}_1,\mathbf{w}_2,\ldots,\mathbf{w}_{N_v}$ based on the span of the top-$N_v$ generalized eigenvectors of $\mathbf{A}$ and $\mathbf{B}$, we will describe a new forward greedy algorithm (i.e., that selects vectors $\mathbf{w}_j$ one at a time without reconsidering past choices) for SIR-maximization  that takes optimal greedy steps at each iteration.  This approach will generally not be worse than the original generalized eigenvalue approach (with respect to the optimization problem from Eq.~\eqref{eq:sir}), and will almost always be better.   

In the first step of the new greedy algorithm, we optimally solve for the first weight vector as
\begin{equation}
  \hat{\mathbf{w}}_{1} = \arg\max_{\mathbf{w} \in \mathbb{C}^{N_v}}  \frac{\mathbf{w}^H\mathbf{A}\mathbf{w} }{\mathbf{w}^H\mathbf{B}\mathbf{w} } \text{ s.t. } \|\mathbf{w}\|_2=1.\label{eq:optim1}
\end{equation}
This optimization problem has the same form as Eq.~\eqref{eq:optim}, and has the same optimal solution described above based on the generalized eigenvalue decomposition of $\mathbf{A}$ and $\mathbf{B}$.  In all subsequent steps, we add a new vector $\hat{\mathbf{w}}_j$ to an existing set of previously selected vectors $\hat{\mathbf{w}}_1,\ldots,\hat{\mathbf{w}}_{j-1}$ in an optimal (but greedy) fashion, by solving
\begin{equation}
  \hat{\mathbf{w}}_{j} = \arg\max_{\mathbf{w} \in \mathbb{C}^{N_v}}  \frac{\mathbf{w}^H\mathbf{A}\mathbf{w} +  \sum_{i=1}^{j-1} \hat{\mathbf{w}}_i^H\mathbf{A}\hat{\mathbf{w}}_i}{\mathbf{w}^H\mathbf{B}\mathbf{w} + \sum_{i=1}^{j-1}\hat{\mathbf{w}}_i^H\mathbf{B}\hat{\mathbf{w}}_i} \text{ s.t. } \|\mathbf{w}\|_2=1 \text{ and } \mathbf{w}^H\hat{\mathbf{w}}_i = 0 \text{ for } i=1,\ldots,j-1.\label{eq:optim}
\end{equation}
This process will proceed iteratively until we have reached the desired number of virtual coils ($j=N_v$).

It remains to derive the optimal solution to Eq.~\eqref{eq:optim}.  In order to simplify the orthogonality constraints in this problem, we will consider a change of variables that lets us remove the constraints entirely.  Let the columns of the matrix $\mathbf{U} \in \mathbb{C}^{N_c \times (N_c-j+1)}$ form an orthonormal basis for the orthogonal complement to the span of the vectors $\hat{\mathbf{w}}_1,\ldots,\hat{\mathbf{w}}_{j-1}$ (which can easily be obtained via the extended singular value decomposition).  Then any vector $\mathbf{w}$ that is orthogonal to $\hat{\mathbf{w}}_1,\ldots,\hat{\mathbf{w}}_{j-1}$ can be represented as $\mathbf{w} = \mathbf{U} \mathbf{z}$ for some $\mathbf{z}\in \mathbb{C}^{(N_c-j+1)}$, and we also have that $\|\mathbf{w}\|_2 = \|\mathbf{z}\|_2$.   For notational simplicity, let's also define $\alpha \triangleq \sum_{i=1}^{j-1} \hat{\mathbf{w}}_i^H\mathbf{A}\hat{\mathbf{w}}_i$ and $\beta \triangleq \sum_{i=1}^{j-1} \hat{\mathbf{w}}_i^H\mathbf{B}\hat{\mathbf{w}}_i$. Then the optimization problem from Eq.~\eqref{eq:optim} can be reexpressed as
\begin{equation}
\begin{split}
 \hat{\mathbf{z}} &= \arg\max_{\mathbf{z}\in\mathbb{C}^{(N_c-j+1)}} \frac{\mathbf{z}^H \mathbf{U}^H\mathbf{A} \mathbf{U} \mathbf{z}+\alpha}{ \mathbf{z}^H \mathbf{U}^H\mathbf{B} \mathbf{U} \mathbf{z} +\beta } \text{ s.t. } \|\mathbf{z}\|_2=1 \\
 &= \arg\max_{\mathbf{z}\in\mathbb{C}^{(N_c-j+1)}} \frac{ \mathbf{z}^H( \mathbf{U}^H\mathbf{A} \mathbf{U}+ \alpha \mathbf{I} )\mathbf{z} }{\mathbf{z}^H( \mathbf{U}^H\mathbf{B} \mathbf{U} +\beta\mathbf{I} )\mathbf{z} } \text{ s.t. } \|\mathbf{z}\|_2=1 .
 \end{split}
\end{equation}
Based on previous arguments, the optimal solution for $\hat{\mathbf{z}}$ can thus be obtained as the unit-normalized generalized eigenvector corresponding to the largest generalized eigenvalue for the matrices $( \mathbf{U}^H\mathbf{A} \mathbf{U} + \alpha \mathbf{I} )$ and $( \mathbf{U}^H\mathbf{B} \mathbf{U} +\beta\mathbf{I} )$, and the optimal value of $\hat{\mathbf{w}}_{j}$ can thus be obtained as $\hat{\mathbf{w}}_{j} = \mathbf{U} \hat{\mathbf{z}}$.

Building the set of $\hat{\mathbf{u}}_j$ vectors recursively in this way is greedy (the solution $\hat{\mathbf{w}}_j$ at each iteration is globally optimal with respect to Eq.~\eqref{eq:optim} assuming that the values of $\hat{\mathbf{w}}_i$ for $i=1,\ldots,j-1$ are fixed, but the collection of vectors $\hat{\mathbf{w}}_{1},\ldots,\hat{\mathbf{w}}_{j}$ is not necessarily globally-optimal with respect to Eq.~\eqref{eq:sir}), but it will lead to simple calculations that will be better with respect to Eq.~\eqref{eq:sir} than the approach described in the original ROVir paper.\footnote{Note that other algorithms do exist \cite{borloz2011} that will achieve even better results, but these are substantially more computationally expensive to implement, and  we do not expect that the computational effort  will be worth the incremental improvement in SIR in most practical situations.}  Specifically, both the original ROVir approach and the new greedy approach choose the same value of $\hat{\mathbf{w}}_1$, but they will often make different choices for $\hat{\mathbf{w}}_2$, and the new greedy approach will never make a worse choice than the original ROVir approach with respect to the cost function from Eq.~\eqref{eq:sir} because we are finding the optimal solution to the exact SIR-optimization problem we are interested in, rather than inheriting a solution from a potentially different optimization problem (Eq.~\eqref{eq:orig}).  Notably, the original ROVir approach would simply determine $\hat{\mathbf{w}}_2$ based on the second generalized eigenvector of $\mathbf{A}$ and $\mathbf{B}$ (suitably orthonormalized with respect to $\hat{\mathbf{w}}_1$), while the new greedy approach determines a potentially different (and always optimal) solution based on the generalized eigendecomposition of a different set of matrices.

\begin{figure}[t]
\centering
\begin{subfigure}{\textwidth}
\includegraphics[width=\textwidth]{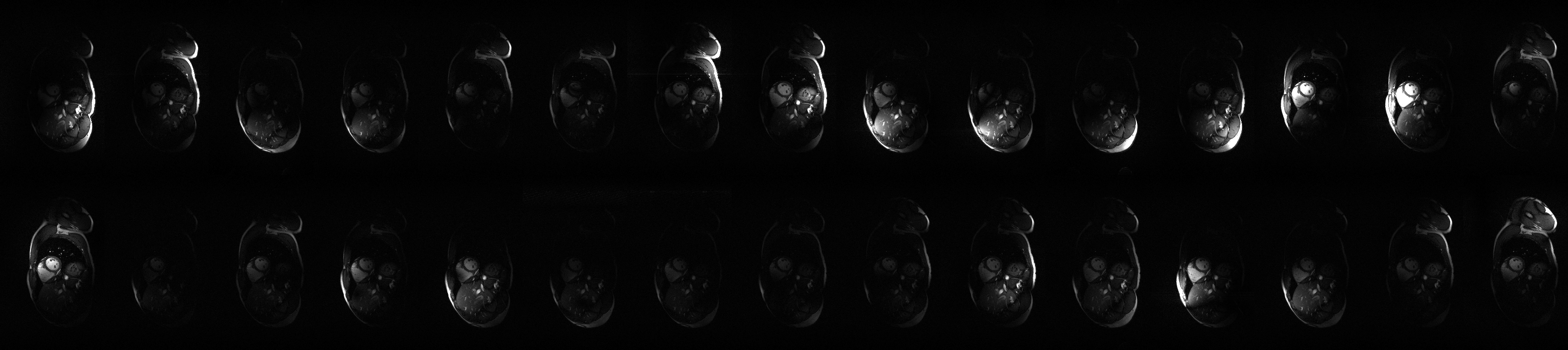}
\caption{Original 30-Channel Dataset}
\end{subfigure}
\begin{subfigure}{0.25\textwidth}
\centering
\includegraphics[width=0.5\textwidth]{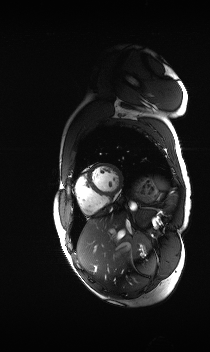}
\caption{Root Sum-of-Squares}
\end{subfigure}
\begin{subfigure}{0.25\textwidth}
\centering
\includegraphics[width=0.5\textwidth]{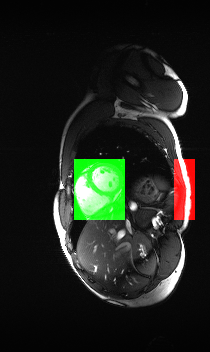}
\caption{ROVir ROIs}
\end{subfigure}
\caption{Cardiac MRI data used for illustration.  In the ROI image, the green region corresponds to the signal region that we want to preserve (which is used to build the matrix $\mathbf{A}$), while the red region corresponds to the interference region that we want to suppress (which is used to build the matrix $\mathbf{B}$).  See Ref.~\cite{kim2021} for additional details.}
\label{fig:case}
\end{figure}
\section{Illustration}
\begin{figure}[t]
\centering
\begin{subfigure}{\textwidth}
\includegraphics[width=\textwidth]{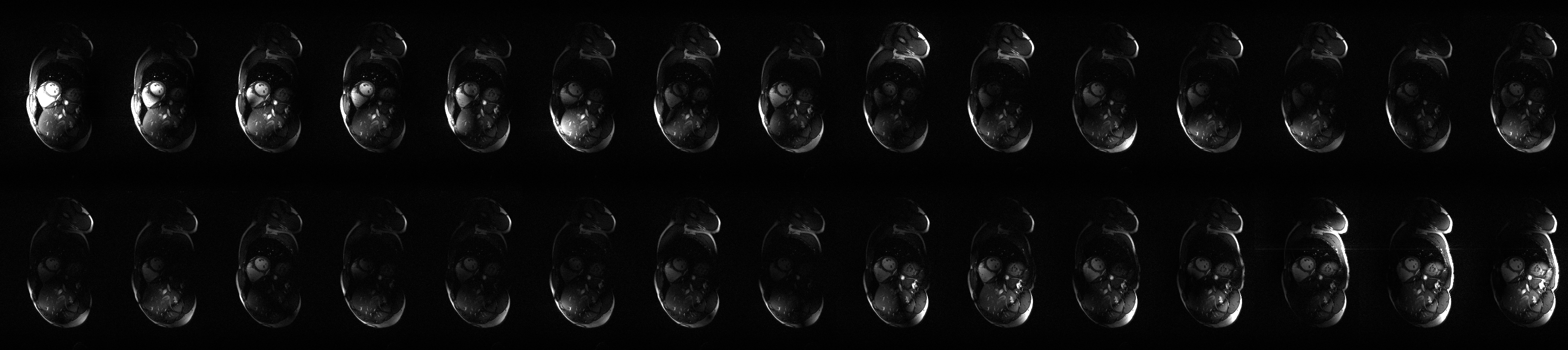}
\caption{Original ROVir Procedure \cite{kim2021}}
\end{subfigure}
\begin{subfigure}{\textwidth}
\includegraphics[width=\textwidth]{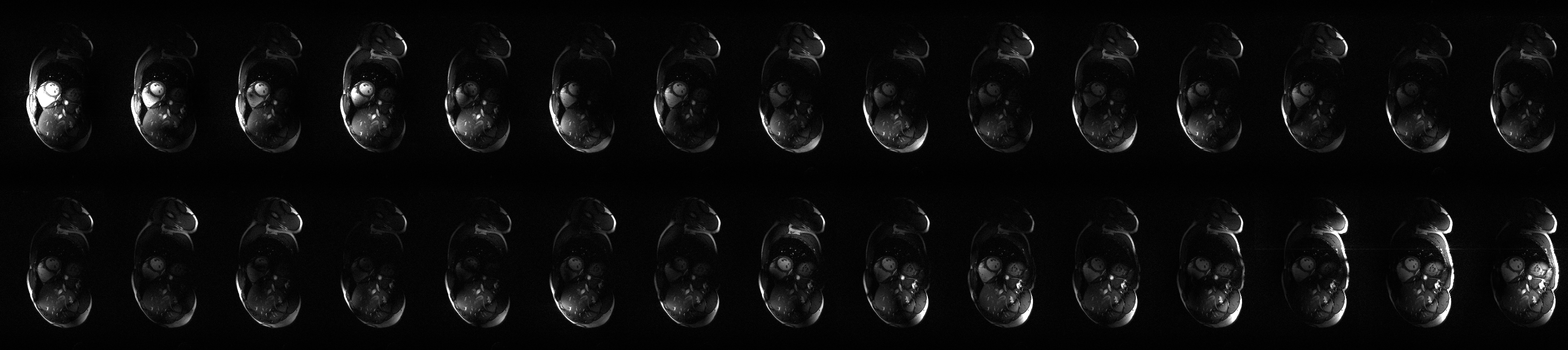}
\caption{New Greedy ROVir Procedure}
\end{subfigure}
\caption{Sets of virtual coils obtained using two different ways of calculating the ROVir coil combination weights.  In each case, virtual coils are shown in order of increasing $j$ (i.e., decreasing SIR) from left-to-right and top-to-bottom.}
\label{fig:virt}
\end{figure}

For illustration, we apply ROVir to a slice from a 30-channel cardiac imaging dataset \cite{kim2023}, as depicted in Fig.~\ref{fig:case}.  We obtained virtual coils from this data using two different approaches: (1) the original ROVir procedure (based on the top-$N_v$ generalized eigenvectors of $\mathbf{A}$ and $\mathbf{B}$) \cite{kim2021}, and (2) the new greedy ROVir procedure introduced in this write-up.  The virtual coils we obtained in each case are depicted in Fig.~\ref{fig:virt}.  As can be seen, both approaches do a reasonably good job of having the energy from the signal region concentrated in the first few virtual coils while having the energy from the interference region concentrated in the last few virtual coils, and the differences between the two calculation approaches are visually quite subtle.

\begin{figure}[t]
\centering
\includegraphics[width=0.3\textwidth]{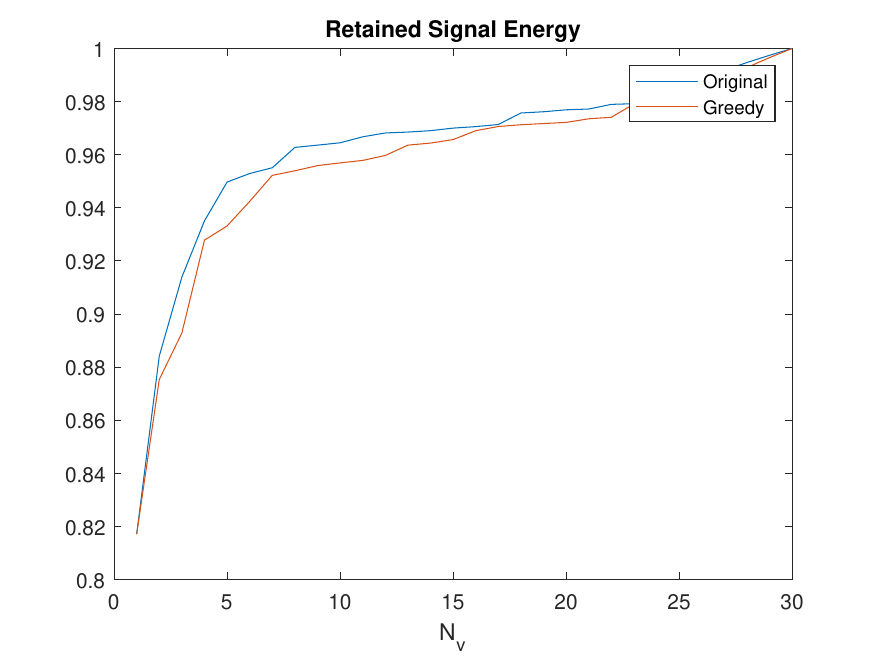}
\includegraphics[width=0.3\textwidth]{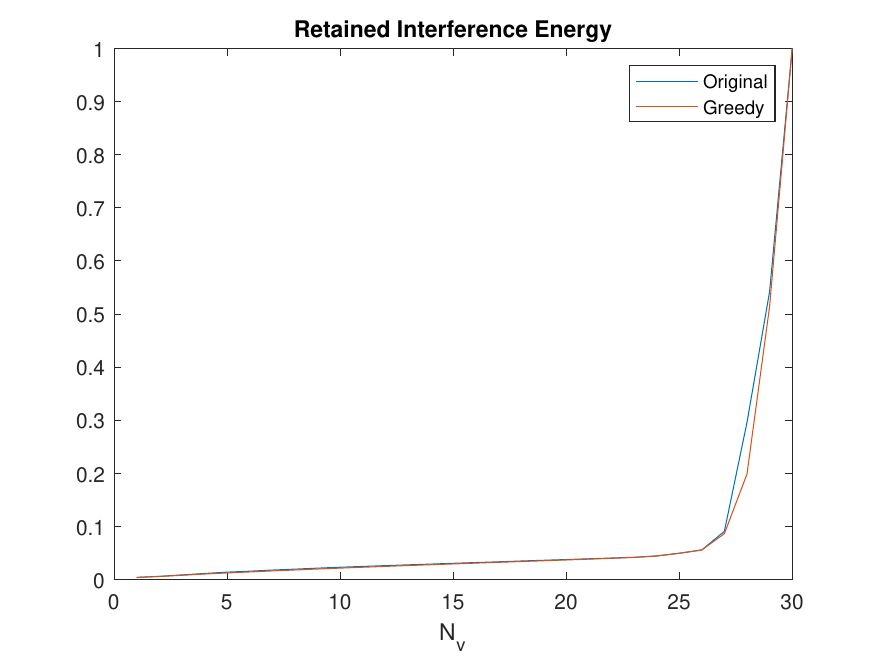}
\includegraphics[width=0.3\textwidth]{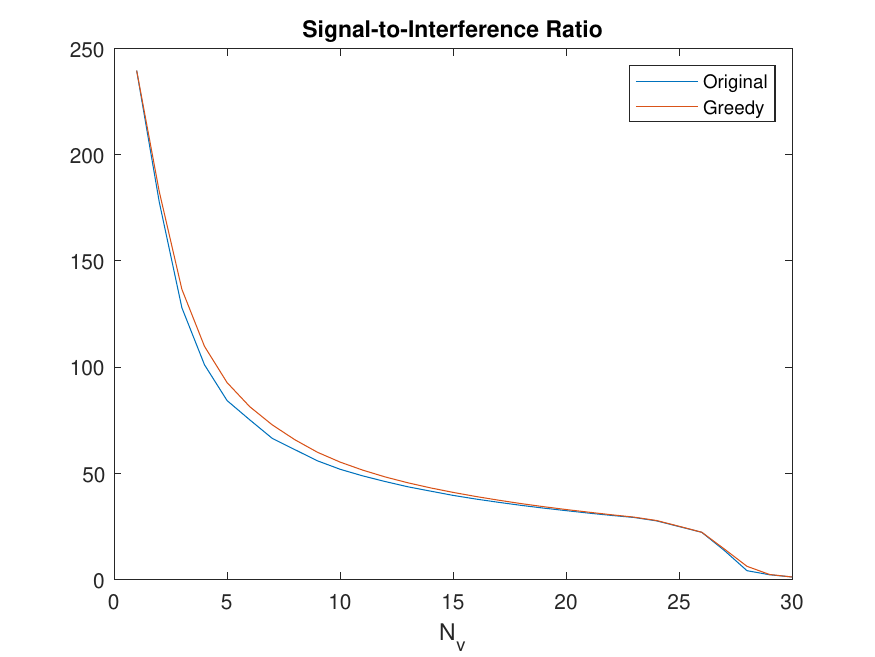}
\caption{Quantitative comparison between the original ROVir procedure and the new greedy ROVir procedure.}
\label{fig:quant}
\end{figure}

\begin{figure}[t]
\centering
\begin{subfigure}{0.25\textwidth}
\centering
\includegraphics[width=0.5\textwidth]{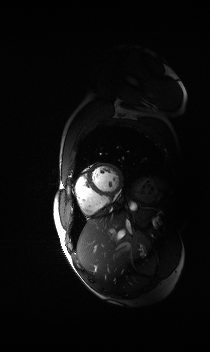}
\caption{Original ROVir}
\end{subfigure}
\begin{subfigure}{0.25\textwidth}
\centering
\includegraphics[width=0.5\textwidth]{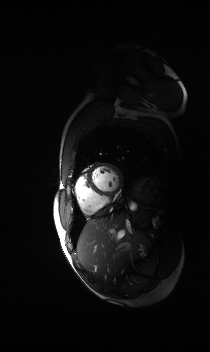}
\caption{Greedy ROVir}
\end{subfigure}
\caption{Results obtained using $N_v=6$ virtual coils.}
\label{fig:rovir} 
\end{figure}

The differences between the two ROVir calculation approaches become more apparent when we start calculating quantitative metrics such as the retained signal energy as a function of $N_v$, the retained interference energy as a function of $N_v$, and the SIR from Eq.~\eqref{eq:sir} as a function of $N_v$.  (See Ref.~\cite{kim2021} for definitions).  These quantitative values are plotted in Fig.~\ref{fig:quant}.  As can be seen, the new greedy ROVir calculation approach has a very minor SIR advantage over the original ROVir calculation approach, as expected theoretically.  In this specific case (which may not generalize), this is achieved by slightly improving interference suppression performance while slightly degrading signal preservation performance.  However, the practical differences between these calculation approaches are not very significant, as the images obtained when performing region-optimized coil compression only have minor differences (see Fig.~\ref{fig:rovir}), and both yield excellent signal preservation and interference suppresion.

\section{Conclusion}
This write-up described some of the mathematical nuances of optimality in ROVir, and introduced a new forward greedy algorithm to calculate ROVir weights that can perform even better in terms of SIR  than the calculation procedure described in the original ROVir paper (at the expense of slightly increased computation).  We believe that this discussion is largely academic -- the original ROVir procedure has already demonstrated excellent performance in a range of different scenarios \cite{kim2021,lam2022,kim2023,schauman2022,iyer2022,iyer2023,schar2023,brackenier2023, piek2023, sorland2023,jin2023,borisch2023 }, and while this new procedure is expected to improve this performance even further, our empirical experience suggests that these improvements may not be very substantial in practice. 

\section*{Acknowledgments}
Thanks to Chin-Cheng Chan for providing comments on a draft of this write-up.

\bibliographystyle{IEEEtran}
\bibliography{./bibliography}

\end{document}